\documentclass{iau} 
\usepackage{graphicx}
\usepackage{natbib}

\title[IAUS 353~~Secular evolution and pseudo-bulges] 
{Secular evolution and pseudo-bulges}

\author[Francoise Combes]   
{Francoise Combes}

\affiliation{Observatoire de Paris, LERMA, Coll\`ege de France, CNRS, \\PSL University, 
Sorbonne University, UPMC, Paris \\ email: {\tt francoise.combes@obspm.fr} }

\pubyear{2019}
\volume{353}  
\jname{Galactic Dynamics in the Era of Large Surveys}
\editors{J. Shen \& M. Valluri}
\begin{document}

\maketitle

\begin{abstract}
Through vertical resonances, bars can produce pseudo-bulges,
within secular evolution. Bulges and pseudo-bulges have doubled their mass since z=1.
The frequency of bulge-less galaxies at z=0 is difficult to explain, especially since
clumpy galaxies at high z should create classical bulges in all galaxies. This issue
is solved in modified gravity models.  Bars and spirals in a galaxy disk,
produce gravity torques that drive the gas to the center and fuel
central star formation and nuclear activity.
 At 0.1-1kpc scale, observations of gravity torques
show that only about one third of Seyfert galaxies experience
molecular inflow and central fueling, while in most cases the
gas is stalled in resonant rings.
At 10-20pc scale, some galaxies have clearly revealed
 AGN fueling due to nuclear trailing spirals,
influenced by the black hole potential.
 Thanks to ALMA, and angular resolution of up to 80mas
 it is possible to reach the central black hole (BH) zone of influence,
 discover molecular tori, circum-nuclear disks misaligned with the galaxy,
and the BH mass can be derived more directly from the kinematics.
\end{abstract}

\firstsection 
\section{Introduction}

The formation of sphero\"idal components has been widely studied through
numerical simulations and observational surveys, and it is possible to
distinguish several scenarios, including secular evolution. There is however
a severe problem still open in the high frequency of bulgeless galaxies at z=0,
which is difficult to account for in the standard model.
Secular evolution through bars is able to form box-peanut shape bulges, and
some are now observed also in nuclear bars. Observations are also able to study
the evolution of pseudo-bulges with redshift.

Bars and embedded structures (nuclear bars and spirals)
are driving gas towards the center of galaxies through gravity torques
and fuel the central black hole. With the recent progress in angular resolution
and sensitivity, we can follow this phenomenon down to the 10pc scales.
It is now possible to detect the molecular tori around nearby Seyfert nuclei,
and to disentangle fueling and molecular outflows.

\begin{figure}[t]
\begin{center}
 \includegraphics[width=0.99\textwidth]{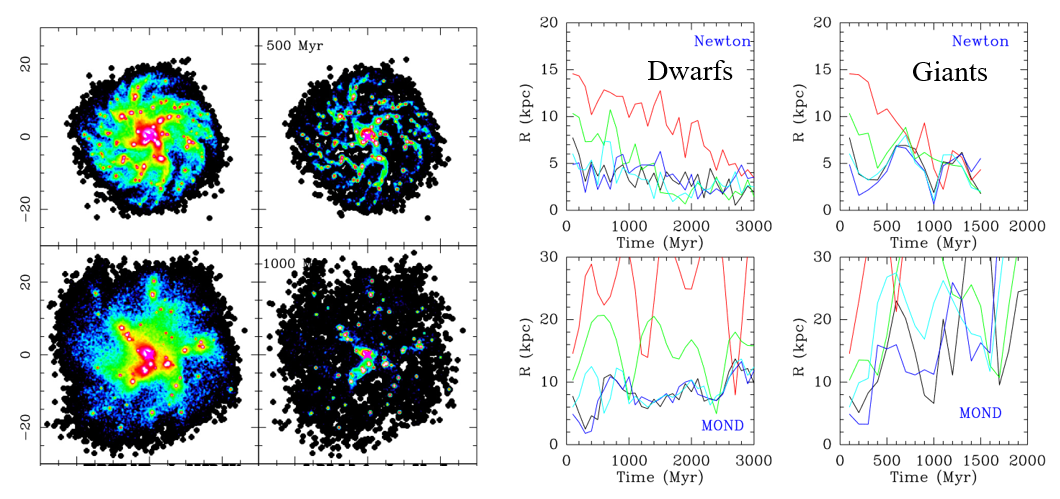} 
\caption{The left panels (each 60kpc in size) correspond to the MOND simulation of a dwarf
galaxy, imaging all  baryons  (left)  and  gas  (right)  surface  densities,
at epochs 0.5, and 1 Gyr. 
 The right panels show the radial decay of the 5 main clumps,	
in the Newtonian+DM (top) and MOND (bottom) for the dwarf (left) and giant (rignt) galaxies
\citep{Combes2014}.}
   \label{fig1}
\end{center}
\end{figure}

\section{The problem of bulgeless galaxies}

Classical bulges and sphero\"ids of early-type galaxies are mostly formed
in mergers, either a small number of major mergers, or a large number
of minor ones, both cancelling out the resulting angular momentum.
Pseudo-bulges are distinct from classical ones, by being more similar to
disks, through their flattening, their rotation level, and their light profile
(Sersic of index n $<$ 3). They are mostly the results of internal
processes, namely vertical resonance with a bar, able to elevate stars in the center
into a thick and sphero\"idal component, more precisely with a peanut-shape
corresponding to the resonance radius.

Galaxies at high redshift are observed clumpy, with 4-5 clumps of at least
10$^9$ M$_\odot$, due to their high gas content and consequent instability.
These clumps can also form a bulge, through dynamical friction on the dark matter halo.
This makes the problem of the high frequency of bulgeless galaxies today 
more acute \citep{Kormendy2008, Weinzirl2009}. Simulations in the standard
model have shown that the supernovae feedback is not rapid enough to destroy
the clumps before their spiraling to the center \citep{Ceverino2010}.

It is however possible to reduce considerably
the dynamical friction, in a modified gravity model, where there is no
dark matter particules, only friction on the disk itself.
Simulations of gaseous galaxies, both dwarfs and giants, have been
carried out with the same star-forming and feedback recipes in MOND and Newton+DM
models \citep{Combes2014}. The clumpy galaxies can evolve into bulgeless
galaxies today (cf Figure \ref{fig1}). In this model, there are also much less galaxy mergers,
so less massive sphero\"ids.

\begin{figure}[t]
\begin{center}
 \includegraphics[width=0.99\textwidth]{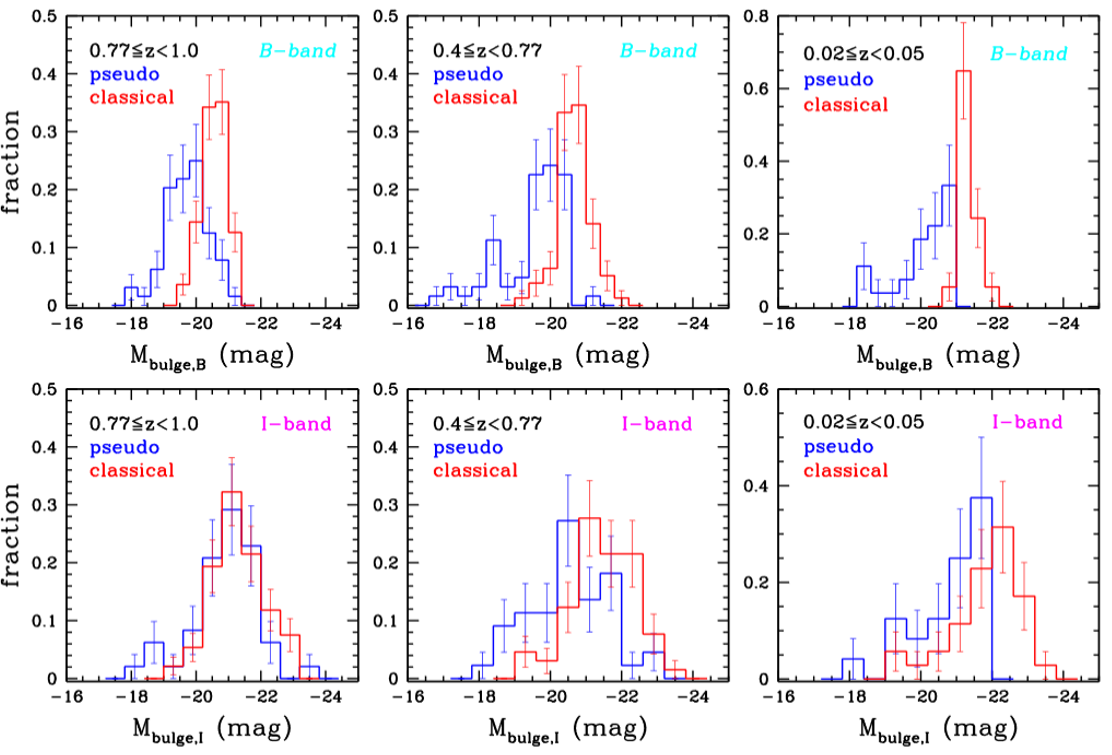} 
	\caption{The distribution of bulge luminosity for pseudo bulges (blue) and for classical 
	bulges (red), is shown for for three redshift ranges, in rest-frame B-band (top)
	and in rest-frame I-band, for HST Good-South (high z) and SDSS (z=0) \citep{Sachdeva2017}.}
   \label{fig2}
\end{center}
\end{figure}

\section{Pseudo-bulges: peanut in nuclear bars, ans evolution with z}

The frequency of pseudo-bulges has been shown to depend
essentially on stellar mass \citep{Fisher2010}. They dominate at
stellar masses lower than 3 10$^{10}$ M$_\odot$, while classical bulges dominate
at larger masses. Their frequency depends also on environment:
there is half less pseudo-bulges in centrals with respect
to satellites and field galaxies \citep{Mishra2017}.
Pseudo-bulges and pure disks  are more frequent  in sheets, and classical bulges
in filaments and halos  \citep{Wang2019}. Small mass galaxies
with classical bulges  have more local neighbors. For them,
intermediate to large-scale environment is not important:
They are associated to neighbors, and are likely due to interactions
and mergers.

The growth of bulges in galaxies since z=1
has been studied by \cite{Sachdeva2017}. The mass in
pseudo-bulges  is about half of the mass of the classical ones.
Both bulges double in mass since z$\sim$1: the bulge to total stellar mass ratio
increases from 10 to 26\% for pseudo-bulge galaxies, and from
21 to 52\% for classical ones. Across redshifts, the growth of bulges
is accompanied by the fading of their disk, which could be attributed
to secular evolution \citep{Sachdeva2017}. Pseudo-bulges are in general less bright
than classical ones, although the difference is fading at high z
in the I-band (cf Figure \ref{fig2}).
The frequency of bars decreases with z, as well as pseudo-bulges
\citep{Melvin2014}.
From a study on HST-COSMOS and SDSS or local galaxies,
\cite{Kruk2018} found that pseudo-bulges start to form
about 7 Gyr ago, at z=0.7-0.8.

Simulations have shown that the vertical rsonance in bars,
responsible to the peanut-shape of their bulges, can alco occur
in nuclear bars. Is it possible to observe them?  Peanut-shape bulges
are easy to recognize in edge-on galaxies, however, if there is an embedded
nuclear bar, it will be difficult to see it in projection on the larger bar.
A nuclear bar is easier to distinguish in more face-on objects.
However, it is possible to recognize a box-peanut bulge even face-on,
through a  barlens \citep{Laurikainen2011, Laurikainen2017}, or through
the h4 parameter in its kinematics \citep{Mendez-Abreu2008, Mendez-Abreu2014}.
The first box-peanut has been detected in a nuclear bar in NGC 1291 by
\cite{Mendez-Abreu2019}.

\begin{figure}[t]
\begin{center}
 \includegraphics[width=0.99\textwidth]{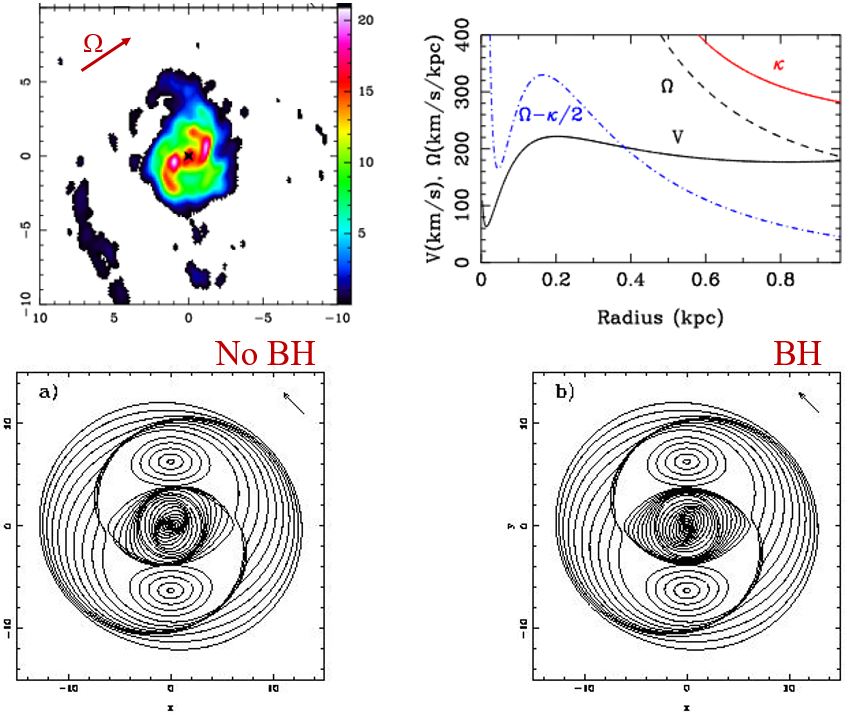} 
	\caption{NGC1566 reveals a trailing nuclear spiral, of ~70pc in radius \citep{Combes+2014}:
	the de-projected map at the top left has been obtained in CO(3-2) with ALMA (axes in arcsec,
	1’’=35pc). The top right panel shows V$_{rot}$, $\Omega$, $\kappa$ and $\Omega -\kappa/2$,
	the precession rate of elliptical orbits. This rate rises again in the sphere of influence
	of the black hole, inside 50pc. Bottom: schematic gas streamlines in the bar rotating frame. 
	The bar is horizontal. With no black hole, a leading spiral is expected,
	which becomes trailing with a black hole, when $\Omega -\kappa/2$ rises again
	\citep{Buta1996}.}
   \label{fig3}
\end{center}
\end{figure}

\section{Fueling and decoupling a secondary bar}

Simulations of the secular evolution of bars have shown how an embedded
nuclear bar can decouple. First, the primary bar grows in amplitude by trapping
more and more orbits, at larger radius, and with a lower precessing rate ($\Omega-\kappa/2$). 
This lowers the pattern speed of the bar $\Omega_b$, which soon develops two
inner Lindblad resonances (ILR). Periodic orbits in between the two ILR are
essentially the x2 family, which is perpendicular to the bar, and 
does not sustain the bar anymore. The bar is further weakened by the vertical
thickening and the formation of the pseudo-bulge \citep{Combes1981}.
Then a new and faster bar develops inside the inner ILR ring \citep{Friedli1993}.

Since bars through their gravity torques drive gas towards the center,
a cool nuclear disk accumulates, forming new stars with low velocity dispersion.
This is at the origin of a drop in dispersion, frequently observed in the center
of nearby Seyfert galaxies, and called $\sigma$-drop. It was observed through near infrared 
spectroscopy by \cite{Emsellem2001}.
The phenomenon was reproduced in simulations by
\cite{Wozniak2003}, taking into account the age of stellar populations, and their
different mass-to-luminosity ratios.
New simulations of $\sigma$-drop \citep{Portaluri2017, diMatteo2019}
have shown that the drop is detected in luminosity weighted lines
until 10 Gyrs.

Gas driven inwards by the bars could fuel an active nucleus. However,
AGN have duty cycles, and the fueling depends on the scales. 
In a previous study of the molecular gas in about 20 nearby Seyferts, with a
spatial resolution of  $\sim$100pc, we conclude that 
only $\sim$35\% of the objects revealed negative torques in the center
\citep{Santi2012}.
The remaining two thirds of galaxies revealed positive torques,
and gas was stalled in a ring at ILR.
Since all galaxies should pass through the fueling phase, this means that 
the fueling phases are short, a few  10$^7$ yrs, and are interrupted by feedback.
Star formation is also simultaneoulsy fueled by the torques, and is
always associated to AGN activity, but can last during longer time-scales.

In the recent years, higher spatial resolution became available with ALMA.
It then was possible to resolve the 10pc scale, and discover the fueling
mechanism closer to the nucleus. For instance, in the case of NGC 1566
(cf Figure \ref{fig3}), in cycle 0 with a beam of 0.5 '' = 25pc, we detected for the
first time a nuclear trailing spiral of R$\sim$ 70pc inside the ILR ring of
radius $\sim$ 420pc \citep{Combes+2014}.
The trailing sense of the spiral was a surprise, since the gravity torques
change sign at each resonance, and around the ILR ring, torques are
negative outside and positive inside, so that gas accumulates in the ring.
This is possible if, at least transiently, gas is in a leading spiral just inside the ring.
But here the nuclear spiral enters in the sphere of influence of the black hole.
The rotation velocity and associated frequencies displayed in
Figure \ref{fig3} show that the $\Omega-\kappa/2$ curve is rising towards
the center, and this explains the reversal of the winding sense of the spiral.
This implies that torques are negative, and the nucleus is being fueled.

\begin{figure}[t]
\begin{center}
 \includegraphics[width=0.99\textwidth]{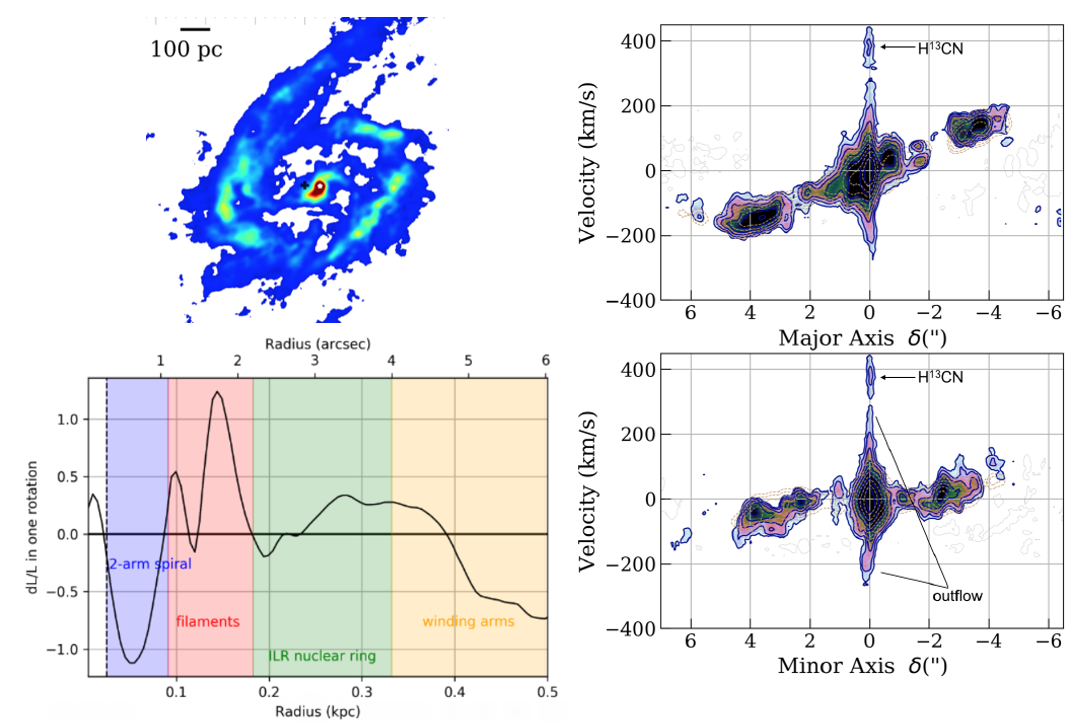} 
	\caption{ALMA CO(3-2) map of the NGC 613 center, revealing a trailing nuclear spiral inside
	the ILR ring. At the bottom left, the gravity torques are negative inside 100pc, gas loses
	all its angular momentum in one rotation. At right, the PV diagrams on major and minor axes
	clearly show a molecular outflow \citep{Audibert2019}.}
   \label{fig4}
\end{center}
\end{figure}

\section{Feedback, outflows, misalignment}

A nuclear trailing spiral has also been detected in NGC 1808
and NGC 613 \citep{Combes2019, Audibert2019}. Although
a molecular outflow has been detected at larger scale in NGC1808
\citep{Busch2017}, there is no outflow in CO at small scale, 
close to the center, and we conclude that the outflow is rather
due to the starburst feedback.
The trailing nuclear spiral has a radius of 45pc, and was observed with a beam
of 0.08'' = 4pc \citep{Audibert2019}. The trailing spiral is also remarkable in the dense
tracers maps: HCN(4-3), HCO+(4-3), CS(7-6).

The case of NGC 613 is even more interesting, revealing both
fueling due to a trailing nuclear spiral, but in simultaneity, a molecular
outflow due to the AGN feedback \citep{Audibert2019}.
With 0.09’’ x 0.06’’ resolution ($\sim$5pc), the  nuclear spiral is
well resolved, with a radius of 60pc, inside the ILR ring of R=300pc.
Inside the spiral, a torus, or circum-nuclar disk of radius 14pc is well
decoupled and contrasted both in its morphology and kinematics, in
particular in the dense tracers \citep{Combes2019}.

Figure \ref{fig4} shows that the gas inside 100pc is losing quickly its
angular momentum. The position-velocity diagrams along the major
and minor axes reveal a molecular outflow \citep{Audibert2019}.
The mass involved in the outflow is 2 10$^6$ M$_\odot$, and the mass outflow
rate is 27 M$_\odot$/yr. This outflow could be entrained by a radio jet, but given the weak
radio activity of NGC 613, it might be a fossil outflow, reflecting its
past activity.

Out of 8 galaxies observed at high resolution, we detected 7 molecular tori,
or circum-nuclear disks. Most of them have misaligned orientation
and random inclinations with respect to the main spiral disk. This is not
unexpected, given the very small scales sampled, and the very different
time-scales involved. These tori are located within the sphere of influence of
their black holes, and see an almost kepkerian spherical potential, with little
influence of the outer disk. They might keep souvenir of their angular momentum,
but the latter could be lost also through star formation feedback and
associated fountain effect.

A high-resolution simulation of a Milky Way-like galaxy
and the zoom in the central 200pc region, 
\citep{Renaud2015, Emsellem2015} have shown how gas can fall back with
any angular momentum. After a starburst episode in the center, a polar
nuclear disk formed, which was rather long-lived, with respect to the
dynamical time \citep{Emsellem2015}.

\section{Summary}

Classical bulges are supposed to form through interactions and
mergers, while pseudo-bulges through internal evolution.
The high frequency of pseudo-bulges or  bulgeless galaxies
among the local universe is a problem for the standard CDM model.
It is possible to reduce the formation of sphero\"ids in a modified gravity model,
where dynamical friction on dark matter halos is suppressed.
 
Bars through their gravity torques drive the gas from the galaxy disk towards the center,
ad can fuel an active nucleus. Fueling phases are short however (a few 10$^7$ yrs), and
at a scale of $\sim$ 100pc, only one third of nearby Seyfert are observed
with negative torques.  The decoupling of embedded nuclear bars or spirals
may take over down to $\sim$ 10pc scales.

With ALMA and improved resolution and sensitivity, it is now possible to resolve these scales,
and several trailing nuclear spirals have been discovered, revealing the fueling
of the AGN at small scales. Molecular tori or circum-nuclear disks
of 10-20 pc in sizes have been detected, randomly oriented with respect
to the main galaxy disk.
In some galaxies, molecular outflows are also observed, either
due to the supernovae feedback or to a radio jet. The coupling of the AGN
feedback is enhanced by the random orientation of the jets.
The misalignment between small and large scales are not unexpected,
especially inside the sphere of influence of the black hole. The gas may
lose its angular momentum orientation, due to stellar feedback
and the fountain effect.


\begin{thebibliography}{}

\bibitem[Audibert et al. (2019)]{Audibert2019}
{Audibert, A., Combes, F., Garcia-Burillo, S. et al.} 2019,
\textit{A\&A}, in press., arXiv1905.01979

\bibitem[Busch et al. (2017)]{Busch2017}
{Busch, G., Eckart, A., Valencia-S, M. et al.} 2017,
\textit{A\&A}, 598, A55

\bibitem[Buta \& Combes (1996)]{Buta1996}
{Buta, R., \& Combes, F.} 1996, 
\textit{ Fund. Cosmic Phys.}, 17, 95

\bibitem[Ceverino et al. (2010)]{Ceverino2010}
{Ceverino, D., Dekel, A., Bournaud, F.} 2010,
\textit{MNRAS}, 404, 2151

\bibitem[Combes \& Sanders (1981)]{Combes1981}
{Combes, F., Sanders, R.H} 1981,
\textit{A\&A}, 96, 164

\bibitem[Combes et al. (2014)]{Combes+2014}
{Combes, F., Garcia-Burillo, S., Casasola, V. et al.} 2014,
\textit{A\&A}, 565, A97

\bibitem[Combes (2014)]{Combes2014}
{Combes, F.} 2014,
\textit{A\&A}, 571, A82

\bibitem[Combes et al. (2019)]{Combes2019}
{Combes, F., Garcia-Burillo, S., Audibert, A et al.} 2019,
\textit{A\&A}, 623, A79

\bibitem[di Matteo et al. (2019)]{diMatteo2019}
{Di Matteo, P., Fragkoudi, F., Khoperskov, S. et al.} 2019,
\textit{A\&A}, 628,A11

\bibitem[Emsellem et al. (2001)]{Emsellem2001}
{Emsellem, E., Greusard, D., Combes, F. et al.} 2001,
\textit{A\&A}, 368, 52

\bibitem[Emsellem et al. (2015)]{Emsellem2015}
{Emsellem, E., Renaud, F., Bournaud, F. et al.} 2015,
\textit{MNRAS}, 446, 2468

\bibitem[Fisher \& Drory (2010)]{Fisher2010}
{Fisher, D. B., Drory, N.} 2010, 
\textit{Ap. J.}, 716, 942

\bibitem[Garcia-Burillo \& Combes (2012)]{Santi2012}
{Garcia-Burillo, S., Combes, F.} 2012, 
\textit{JPhCS}, 372, a2050

\bibitem[Friedli \& Martinet (1993)]{Friedli1993}
{Friedli, D., Martinet, L.} 1993, 
\textit{A\&A}, 277, 27

\bibitem[Kormendy \& Fisher (2008)]{Kormendy2008}
{Kormendy, J., Fisher, D.B.} 2008, 
\textit{ASPC}, 396, 297

\bibitem[Kruk et al. (2018)]{Kruk2018}
{Kruk, S. J., Lintott, C. J., Bamford, S. P. et al.} 2018,
\textit{MNRAS}, 473, 4731

\bibitem[Laurikainen \& Salo (2017)]{Laurikainen2017}
{Laurikainen, E., Salo, H.} 2017,
\textit{A\&A}, 598, A10

\bibitem[Laurikainen et al. (2011)]{Laurikainen2011}
{Laurikainen, E., Salo, H., Buta, R., Knapen, J. H.} 2011,
\textit{MNRAS}, 418, 1452

\bibitem[Melvin et al. (2014)]{Melvin2014}
{Melvin, T., Masters, K., Lintott, C., et al.} 2014,
\textit{MNRAS}, 438, 2882

\bibitem[Mendez-Abreu et al. (2008)]{Mendez-Abreu2008}
{Mendez-Abreu, J., Corsini, E. M., Debattista, V. P. et al.} 2008,
\textit{Ap. J.}, 679, L73

\bibitem[Mendez-Abreu et al. (2014)]{Mendez-Abreu2014}
{Mendez-Abreu, J., Debattista, V. P., Corsini, E. M., Aguerri, J. A. L.} 2014,
\textit{A\&A}, 572, A25

\bibitem[Mendez-Abreu et al. (2019)]{Mendez-Abreu2019}
{Mendez-Abreu, J., de Lorenzo-Caceres, A., Gadotti, D. A. et al.} 2019,
\textit{MNRAS}, 482, L118

\bibitem[Mishra et al. (2017)]{Mishra2017}
{Mishra, P. K., Wadadekar, Y., Barway, S.} 2017,
\textit{MNRAS}, 467, 2384

\bibitem[Portaluri et al. (2017)]{Portaluri2017}
{Portaluri, E., Debattista, V. P., Fabricius, M. et al.} 2017,
\textit{MNRAS}, 467, 1008

\bibitem[Renaud et al. (2015)]{Renaud2015}
{Renaud, F., Bournaud, F., Emsellem, E. et al.} 2015,
\textit{MNRAS}, 454, 3299

\bibitem[Sachdeva et al. (2017)]{Sachdeva2017}
{Sachdeva, S., Saha, K., Singh, H. P.} 2017,
\textit{Ap. J.}, 840, 79 

\bibitem[Wang et al. (2019)]{Wang2019}
{Wang, L., Wang, L., Li, C. et al.} 2019,
\textit{MNRAS}, 484, 3865

\bibitem[Weinzirl et al. (2009)]{Weinzirl2009}
{Weinzirl, T., Jogee, S., Khochfar, S. et al.} 2009,
\textit{Ap. J.}, 696, 411

\bibitem[Wozniak et al. (2003)]{Wozniak2003}
{Wozniak, H., Combes, F., Emsellem, E., Friedli, D.} 2003,
\textit{A\&A}, 409, 469

\end{thebibliography}
\end{document}